\documentclass[twocolumn,
aps,nofootinbib,showpacs,showkeys,preprint
tightenlines,preprintnumbers,] {revtex4}

\usepackage{epsf,epsfig,subfigure,graphicx,amsmath,amssymb ,mathtools
}
\usepackage{color}
\newcommand{\dis}[1]{\begin{equation}\begin{split}#1\end{split}\end{equation}}
\newcommand{\ie}{{\it i.e.~}}
\newcommand{\etal}{{\it et al.\,}}

\newcommand{\gev}{\,\textrm{GeV}}

\newcommand{\UULA}{U(1)$_{\rm ULA}$}
\newcommand{\UPQ}{U(1)$_{\rm PQ}$}
\newcommand{\be}{\begin{equation}}
\newcommand{\ee}{\end{equation}}
 
\def\sw0{{$\sin^2\theta_W^0$}}

\def\ula{\phi}
\def\fula{f_{\rm ULA}}

\newcommand{\Z}{{\bf Z}}

\def\E6{{\rm E_6}}

\def\EE8{{\rm E_8\times E_8'}}

\def\fqcd{f_{\rm QCD}}
\def\fula{f_{\rm ULA}}
\def\mula{m_{\rm ULA}}
\def\tqcd{\theta_{\rm QCD}}
\def\tula{\theta_{\rm ULA}}

\newcommand{\jcap}{JCAP}		
\newcommand{\jhep}{JHEP}		

\begin{document}

\title{An ultralight pseudoscalar boson}

\author{Jihn E. Kim,$^{(a,b)}$ David J. E. Marsh$^{(c,d)}$ }
 
\affiliation{
 $^{(a)\,}$Department of Physics, Kyung Hee University, 26 Gyungheedaero, Dongdaemun-Gu, Seoul 02447, Republic of Korea,  \\
$^{(b)\,}$Center for Axion and Precision Physics Research (IBS),
  291 Daehakro, Yuseong-Gu, Daejeon 34141, Republic of Korea,\\ 
$^{(c)\,}$Perimeter Institute, 31 Caroline St N,  Waterloo, ON, N2L 6B9, Canada \\
$^{(d)\,}$ Department of Physics, King's College London, Strand, London, WC2R 2LS, United Kingdom
}

\begin{abstract} 
Using a fundamental discrete symmetry, $\Z_N$, we construct a two-axion model with the QCD axion solving the strong-$CP$ problem, and an ultralight axion (ULA) with $m_{\rm ULA}\approx 10^{-22}\text{ eV}$ providing the dominant form of dark matter (DM). The ULA is light enough to be detectable in cosmology from its imprints on structure formation, and may resolve the small-scale problems of cold DM. The necessary relative DM abundances occur without fine tuning in constructions with decay constants $\fula\sim 10^{17}\text{ GeV}$, and $\fqcd\sim 10^{11}\text{ GeV}$. An example model achieving this has $N=24$, and we construct a range of other possibilities. We compute the ULA couplings to the Standard Model, and discuss prospects for direct detection. The QCD axion may be detectable in standard experiments through the $\vec{E}\cdot\vec{B}$ and $G\tilde{G}$ couplings. In the simplest models, however, the ULA has identically zero coupling to both $G\tilde{G}$ of QCD and $\vec{E}\cdot\vec{B}$ of electromagnetism due to vanishing electromagnetic and color anomalies. The ULA couples to fermions with strength $g\propto 1/\fula$. This coupling causes spin precession of nucleons and electrons with respect to the DM wind with period $t\sim$months. Current limits do not exclude the predicted coupling strength, and our model is within reach of the CASPEr-Wind experiment, using nuclear magnetic resonance.  
\end{abstract}
\preprint{KCL-PH-TH/2015-44}
\maketitle


\section{Introduction}\label{sec:Introduction}

Dark matter (DM) is known to comprise around 30\% of the energy density of the Universe~\cite{Planck15}, yet its origin within a fundamental theory remains unknown. A classic candidate embedded in a minimal extension of the Standard Model (SM) is the QCD axion~\cite{Kim79,SVZ80,Zhit80,DFS81} based on the Peccei-Quinn (PQ) symmetry \cite{PQ77,Weinberg78,Wilczek78,Baer15}. In addition to being a good DM model, the QCD axion also solves the strong-$CP$ problem, by setting the (static) neutron electric dipole moment to zero, consistent with observations~\cite{NEDMexp06}. The QCD axion behaves, for all intents and purposes, like cold dark matter (CDM): its background energy density scales like $a^{-3}$ (where $a$ is the cosmic scale factor) on all post-big bang nucleosynthesis time scales, and the Jeans scale associated to the gradient energy is vanishingly small on astrophysical length scales~\cite{Khlopov85}. However, CDM, and thus the QCD axion, suffers from some possible problems stemming from exactly these properties: the scale invariance of CDM structure formation leads to the `small-scale crises'~\cite{SmallCrisis06}. 

The small-scale crises may be alleviated by proper inclusion of the effects of star formation and feedback (see e.g. Ref.~\cite{SmallFeedback} for recent simulations, and Ref.~\cite{SmallRev} for a review). The small-scale crises can also be addressed by allowing more freedom in the model of DM, deviating from exact CDM. One possibility is to introduce thermal velocities, so-called warm dark matter (WDM)~\cite{Bond82,Bode2001}, and particle physics candidates include the gravitino and sterile neutrinos. WDM, however, cannot give large cores to dwarf spheroidal (dSph) galaxies while also being consistent with cosmological structure formation~\cite{MNRAS12}, i.e. WDM cannot completely solve the small-scale crises. 

In the context of axion theories, an ultralight axion (ULA) offers an elegant solution~\cite{Hu00,MarshSilk14}. ULAs differ from CDM essentially because of the large de Broglie wavelength, which imprints a scale on structure formation, suppressing linear density perturbations (e.g. Refs.~\cite{MarshFerr14,Park12}) and allowing for the formation of cored pseudo-solitons on non-linear scales~\cite{Ruffini69,Seidel91,SchiveNat14,SchivePRL14}. A ULA with mass $m_{\rm ULA}\approx 10^{-22}\text{ eV}$ making up a dominant ($\gtrsim 90\%$) component of the DM can provide large cores to dSph galaxies~\cite{MarshPop15} (solving the `cusp-core' problem) and is consistent with both the CMB~\cite{Hlozek15} and high-$z$ galaxy formation~\cite{Bozek15,SchiveHighZ}, thus avoiding the \emph{Catch 22} of WDM. If the DM is dominantly composed of a ULA with $10^{-22}\text{ eV}\lesssim m_{\rm ULA}\lesssim 10^{-18}\text{ eV}$, then irrespective of its role in the small-scale crises its effects may be detectable in the epoch of reionization~\cite{Bozek15,SchiveHighZ} and the 21cm power spectrum~\cite{MarshPRD15}. The origin of such a small mass scale for the ULA, which is at the same time able to give a dominant contribution to the DM, is the subject of this paper.

In this paper, we aim to construct a model based on principles dictated by quantum gravity in which such a DM model is somewhat natural. We use a fundamental discrete symmetry $\Z_N$ giving rise to approximate global symmetries. We demand the existence of the QCD axion and a successful resolution of the strong-$CP$ problem: by symmetry arguments this dictates the couplings of the two-axion model. The QCD axion should, however, be only a fraction of the DM. This occurs with minimal fine-tuning and model uncertainty if the axion decay constant $\fqcd\lesssim 10^{14}\text{ GeV}$, and the PQ symmetry, \UPQ, is broken before or during inflation. The rest of the DM is composed of a ULA with $m_{\rm ULA}\approx 10^{-22}\text{ eV}$, whose mass is protected from large quantum corrections by the discrete symmetry. In order for this ULA to contribute a large amount to the DM density its decay constant must be $f_{\rm ULA}\sim 10^{17}\text{ GeV}$. The corresponding PQ symmetry, \UULA, must also be broken before or during inflation. Constraints from isocurvature perturbations require that the Hubble scale during inflation be $H_I< 10^{10}\text{ GeV}$, and so this model predicts vanishingly small tensor perturbations, $r_T\ll 10^{-9}$~\cite{MarshPRD13,MarshPRL14}. The demands on $m_{\rm ULA}$ and $f_{\rm ULA}$ place restrictions on the possible symmetry groups $\Z_N$.

The benefit of constructing an explicit QCD-ULA model is that we can compute all of the couplings of the ULA to the SM, which has not been possible before. Demanding a successful solution to the strong-$CP$ problem puts constraints on these couplings, and in the simplest constructions forbids a ULA coupling to $\vec{E}\cdot\vec{B}$ of electromagnetism, thus making it invisible to many standard searches for axion-like particles (ALPs, e.g. Ref.~\cite{Dias14}, which also discusses the role of discrete symmetries). The demand of vanishing ULA color anomaly does, however, predict tree-level couplings between the ULA and SM fermions, which may be detectable via searches for spin precession of nucleons~\cite{Graham13,casper}. These are key results of this paper. 

The existence of ULAs is often discussed within a string theory context~\cite{Witten84,Svrcek06,Arvanitaki10,KimPRL13}. Our construction is based purely on field theory,\footnote{An alternative field theory model is discussed in Ref.~\cite{Chiueh14}.} though the use of discrete symmetries is consistent with expectations about global symmetries in quantum gravity (e.g. Refs.~\cite{Krauss89,Kamion92,BarrBlock}). Discrete symmetries may arise in phenomenologically consistent orbifold compactifications of string theory \cite{KimPLB13,KimNilles14,KimJKPS14}. 

The rest of the paper is organized as follows. In Section~\ref{sec:dm_model} we describe our criteria for the DM model. In Section~\ref{sec:discrete_symms} we build the desired model, and a number of variants. We compute the axion couplings to the SM in Section~\ref{sec:couplings}, and analyze the direct and indirect detection possibilities in Section~\ref{sec:detection}. We conclude in Section~\ref{sec:conclusions}. In the Appendix we present another model variant with additional heavy quarks and a non-vanishing ULA coupling to electromagnetism.

\section{Two-Component Axion Dark Matter}
\label{sec:dm_model}

Our model contains two axions: the QCD axion with decay constant $\fqcd$ arising from a spontaneously broken symmetry \UPQ, and a ULA with decay constant $\fula$ and mass $\mula$ arising from a spontaneously broken symmetry \UULA. We impose a constraint on the QCD axion decay constant $10^9\text{ GeV}<\fqcd<10^{17}\text{ GeV}$, where the lower bound comes from e.g. supernova cooling~\cite{Grifols96,Raffelt96}, and the upper bound from black hole superradiance~\cite{Lyth92}. We are interested in ULA models with observable cosmological consequences, and limit the ULA mass to be in the range $10^{-22}\text{ eV}<\mula<10^{-18}\text{ eV}$ where the lower limit is allowed by cosmic structure formation~\cite{Hlozek15,Bozek15,SchiveHighZ} and the upper limit is the possible reach of 21cm surveys~\cite{MarshPRD15}. Other relevant cosmological parameters are listed in Table~\ref{tab:cosmo_params}.

When \UPQ~and \UULA~are both broken before or during inflation, $f_{\rm X}>H_I/2\pi$ ($H_I$ is the Hubble scale during inflation; we discuss the reason for this assumption shortly), the dominant axion production is non-thermal, occuring via vacuum realigment with the axions at essentially zero temperature. The relic density in each axion, X, is determined by the spatially averaged misalignment angle:
\be
\langle \theta_{i,\rm X}^2\rangle=\bar{\theta}_{i,\rm X}^2+(H_I/2\pi f_{\rm X})^2\, .
\label{eqn:average_mis}
\ee

The first term in Eq.~\eqref{eqn:average_mis} is the initial random value from spontaneous symmetry breaking smoothed by inflation, and the second term arises from back-reaction of the inflationary isocurvature perturbations~\cite{Lyth92}. The relic densities from vacuum realignment are:\footnote{These are approximate expressions, but are accurate enough for the purposes of the present study. For more details, see e.g. Refs.~\cite{Fox04,Hlozek15,Wantz09,Visinelli09}.}
\begin{align}
\Omega_{\rm QCD}h^2&=2\times 10^4\left(\frac{\fqcd}{10^{16}\text{GeV}}\right)^{7/6}\langle \theta_{i,\rm QCD}^2\rangle \mathcal{F}_{\rm anh.}(\bar{\theta}_{i,\rm QCD})\, ,\label{eqn:relic_qcd} \\
\Omega_{\rm ULA}h^2&=\frac{h^2}{6}(9\Omega_r)^{3/4}\left(\frac{\mula}{H_0}\right)^{1/2}\left(\frac{\fula}{M_{pl}}\right)^2\langle(\theta_{i,\rm ULA})^2\rangle \, \label{eqn:relic_ula},
\end{align}
where $h$ is the reduced Hubble rate today, $H_0=100h\text{ km s}^{-1}\text{Mpc}^{-1}$, and $\Omega_r$ is the energy density in radiation, determined by the redshift of equality, $z_{\rm eq}$, and the total matter density, both of which we hold fixed. The factor $\mathcal{F}_{\rm anh.}(x)$ accounts for anharmonic corrections to the QCD axion relic density, assuming a cosine potential, $V(\tqcd)\propto (1-\cos \tqcd)$, for which we use the analytic fit of Ref.~\cite{Visinelli09}. We only assume a quadratic ULA potential, $V(\tula)\propto\tula^2$, which is valid for $\tula\lesssim 1$. The ULA relic density depends on the combination of parameters $\bar{\theta}_{i,\rm ULA}\fula\equiv\phi_{i}$, and thus the relic density in each axion is a function of three free parameters: $\Omega_{\rm QCD}=\Omega_{\rm QCD}(f_{\rm QCD},\bar{\theta}_{i,\rm QCD},H_I)$, $\Omega_{\rm ULA}=\Omega_{\rm ULA}(m_{\rm ULA}, \phi_i,H_I)$ .

The QCD relic density, Eq.~\eqref{eqn:relic_qcd}, is strictly valid only for $\fqcd\lesssim 2\times 10^{15}\text{ GeV}$~\cite{Fox04,Wantz09}. The differences at large $\fqcd$ are fairly small on the scale of the present analysis, and in most cases we restrict ourselves to smaller $\fqcd$ anyway. The ULA relic density, Eq.~\eqref{eqn:relic_ula}, is valid as long as ULA coherent oscillations begin in the radiation dominated era. Oscillations begin when $H\sim \mula$, and the Hubble rate at $z_{\rm eq}$ is $H_{\rm eq}\sim 10^{-28}\text{ eV}$. We are concerned only with ULAs that are allowed to be the dominant form of DM, which requires at the very least $m>10^{-24}\text{ eV}$~\cite{Hlozek15}. Eq.~\eqref{eqn:relic_ula} is thus an excellent approximation. We assume that temperature dependence of the ULA mass can be neglected, or more accurately that the mass has reached its zero-temperature value by the time coherent oscillations begin.\footnote{See Ref.~\cite{Dienes15} for an interesting discussion of the role of mixing and mass evolution in multi-axion systems.}
\begin{figure*}[!ht]
\begin{center}
\begin{tabular}{cc}
 \includegraphics[width=0.5\textwidth]{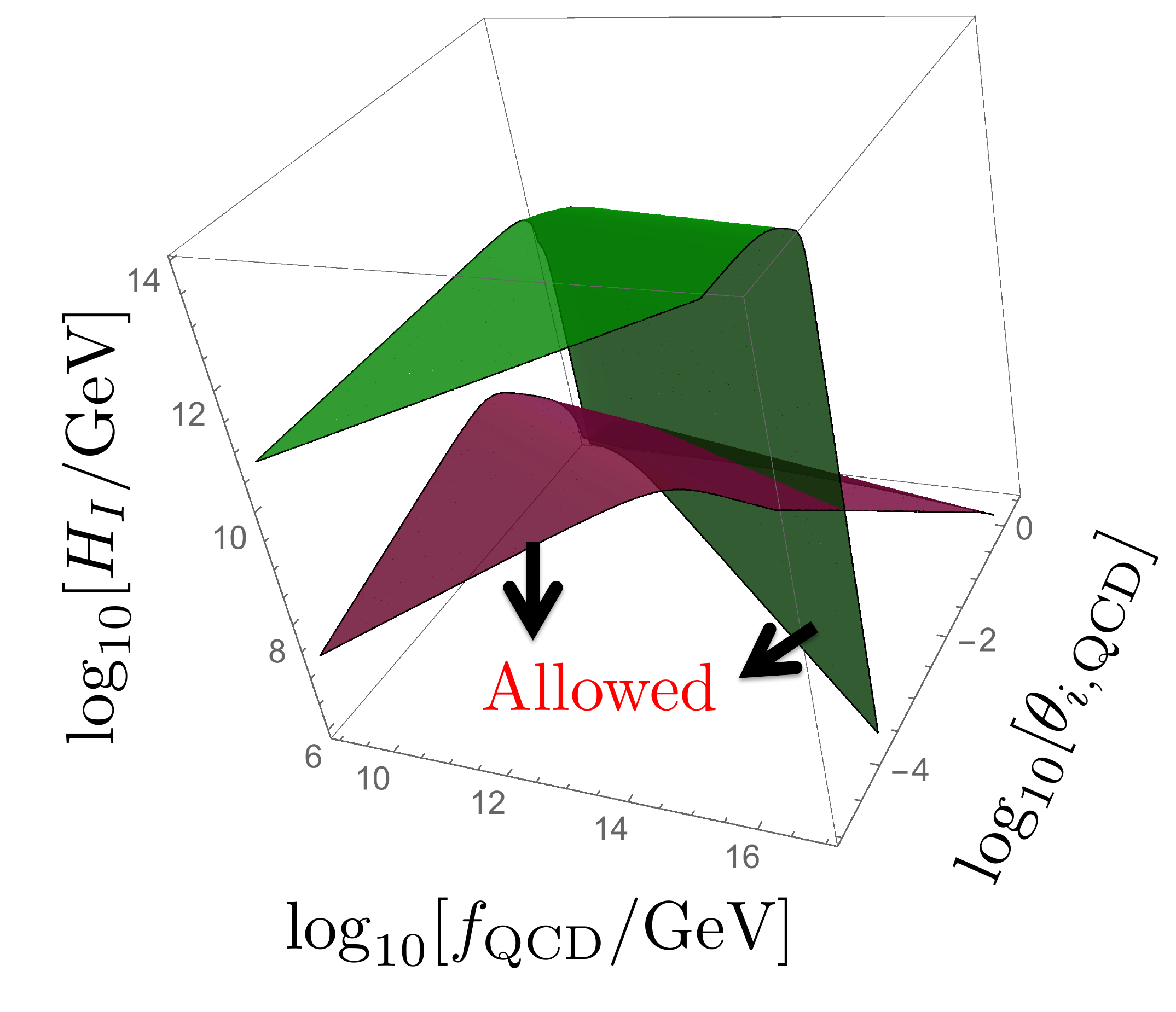}
\includegraphics[width=0.5\textwidth]{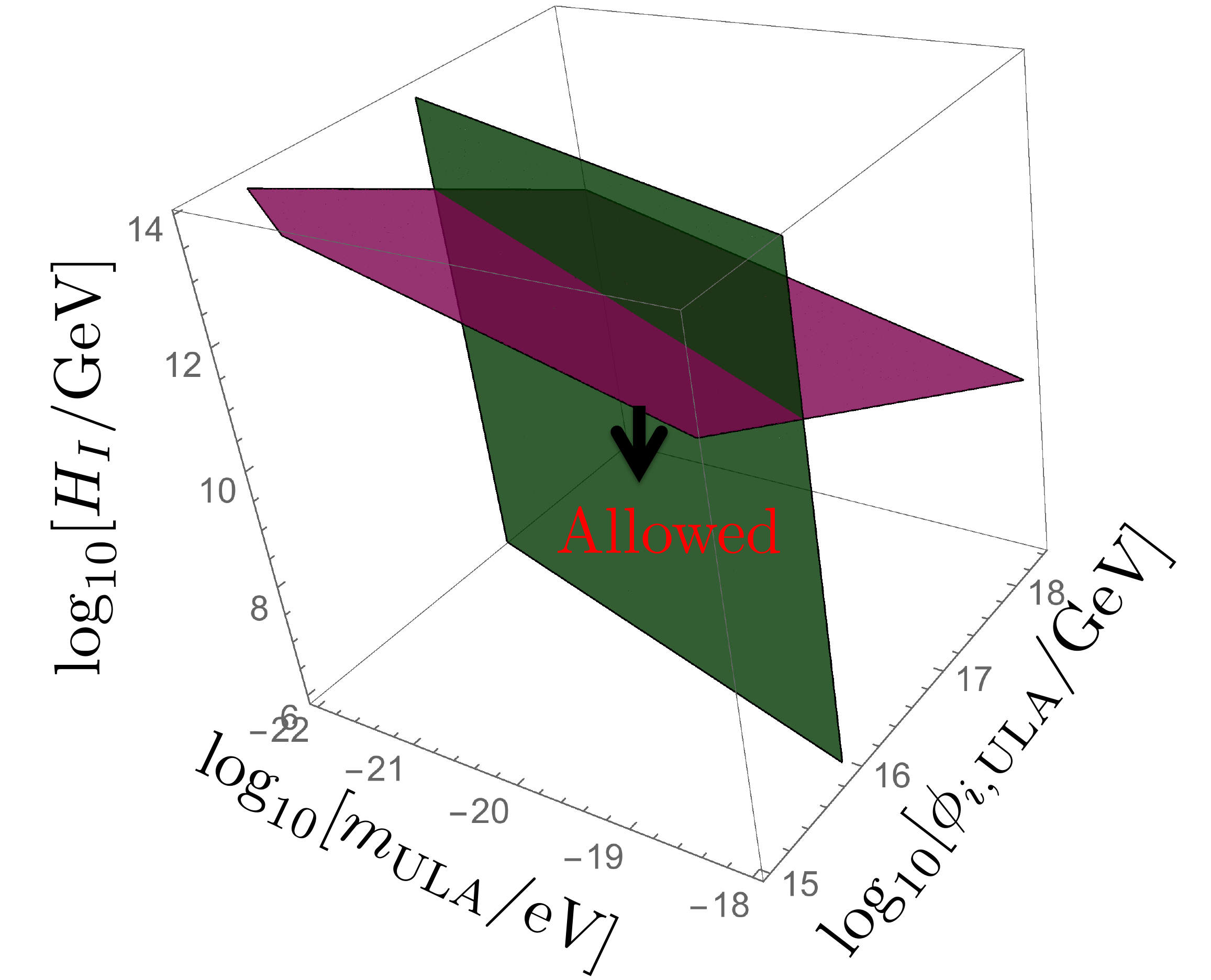}
 \end{tabular}
 \end{center}
\caption{Constraints on the QCD axion and ULA parameter space imposed by DM relic density (green) and isocurvature (purple) in the scenario where \UPQ~and \UULA~are both broken during inflation. \emph{Left panel}: QCD axion. We demand this is a sub-dominant DM component $\Omega_{\rm QCD}/\Omega_c<0.1$. The maximum allowed inflationary scale is $H_I\sim 10^{10}\text{ GeV}$, with the full parameter space consistent with $H_I\lesssim 10^8\text{ GeV}$. All $\fqcd<10^{17}\text{ GeV}$ can be accommodated with $\bar{\theta}_{i,\rm QCD}\gtrsim 10^{-4}$ fine tuning. Restricting to $\bar{\theta}_{i,\rm QCD}\geq 10^{-2}$ imposes $\fqcd\lesssim 10^{14}\text{ GeV}$. \emph{Right panel:} ULA. For the range of masses and relic densities considered, backreaction of isocurvature perturbations is negligible and the relic density is independent of $H_I$ for $r_T<0.12$. Isocurvature perturbations constrain $H_I\lesssim 10^{12}\text{ GeV}$. On the scale shown, obtaining relic density $0.9<\Omega_{\rm ULA}/\Omega_c<1$ constrains the model to live in the green plane, implying $\fula>10^{16}\text{ GeV}$ if $\bar{\theta}_{i,\rm ULA}< 1$.}
\label{fig:surfaces}
\end{figure*}

\begin{table}
\centering
\begin{tabular}{c|c|c}
\hline \hline
Parameter&Constraint&Ref.
 \\ \hline \hline
$\Omega_c h^2$&$0.1193\pm 0.0014$&\cite{Planck15}  \\ 
$z_{\rm eq}$& $3382\pm 32$ &\cite{Planck15} \\
$h$&$0.6751\pm0.0064 $& \cite{Planck15} \\
$\Omega_m$& $0.3121\pm 0.0087$& \cite{Planck15} \\
$\ln (10^{10}A_s)$& $3.059\pm 0.025$ & \cite{Planck15} \\
$A_I/A_s$& $<0.038$& \cite{Planck15Inflation} \\
$r_T$& $<0.12$ & \cite{BKPlanck15} \\
\hline\hline
\end{tabular}
\caption{Relevant cosmological parameters, as defined in the text. Errors are 68\% C.L., upper limits are 95\% C.L. }
 \label{tab:cosmo_params}
\end{table}

When the PQ symmetry is broken during inflation the axion fields obtain (uncorrelated) isocurvature perturbations with amplitude
\be
A_{I,\rm X}=\left(\frac{\Omega_{\rm X}h^2}{\Omega_c h^2}\right)^2\left(\frac{H_I}{\pi\bar{\theta}_{i,\rm X}f_{\rm X}}\right)^2\, .
\ee
The total DM density is $\Omega_c h^2=\Omega_{\rm QCD}h^2+\Omega_{\rm ULA}h^2$ and the total isocurvature amplitude is $A_I=A_{I,\rm QCD}+A_{I,\rm ULA}$. For the ULA masses we consider, both the QCD axion and the ULA are indistinguishable from CDM in terms of CMB data~\cite{Hlozek15,MarshPRD13,MarshPRL14}, and so only these total combinations, $\{\Omega_c h^2, A_I\}$, are constrained. The value of $H_I$ is bounded from above by the combined \emph{Planck}/BICEP2 constraint~\cite{BKPlanck15} on the tensor-to-scalar ratio, $r_T=(8/A_s)(H_I/2\pi M_{pl})^2$, where $M_{pl}=1/\sqrt{8\pi G_N}$ is the reduced Planck mass. 

We are interested in models where the DM is dominantly composed of ULAs, with a small component of the QCD axion, and in which this can be obtained ``naturally,'' i.e. without too much fine-tuning. We demand that $\Omega_{\rm QCD}/\Omega_c<0.1$ with the rest of the DM in ULAs. We impose naturalness by limiting the misalignment angles $\bar{\theta}_{i,\rm X}>\bar{\theta}_{i,\rm min}$, with $\bar{\theta}_{i,\rm min}$ depending on the severity of fine-tuning we are prepared to allow (recall that the strong-$CP$ problem itself, which we are trying to alleviate by introducing the QCD axion, is a tuning of $\theta_{\rm QCD}\sim 10^{-10}$). 

Fig.~\ref{fig:surfaces} sketches the constraints on our model imposed by the DM relic density and isocurvature bounds.\footnote{Since the chosen ULA masses behave like CDM in the CMB, it is a trivial matter to perform a full analysis from the public \emph{Planck} chains, using the given formulae and imposing our priors. At the present level of model-building, this level of detail is unnecessary and the sketched constraints are a good enough guide on the logarithmic scale of parameters.} The dominant ULA component of DM requires $\phi_i>10^{16}\text{ GeV}$ and thus $\fula>10^{16}\text{ GeV}$ for $\tula<1$. The constraint $r_T<0.12\Rightarrow H_I<8.8\times 10^{13}\text{ GeV}$ implies that ULAs must have \UULA~broken during inflation. Isocurvature constraints on the ULAs then further limit $H_I\lesssim 10^{12}\text{ GeV}$, implying all $\fqcd\gtrsim 10^{11}\text{ GeV}$ in our model have \UPQ~broken during inflation. The strongest constraint on $H_I$ comes from the QCD axion, which even as a subdominant DM component demands $H_I\lesssim 10^{10}\text{ GeV}$, falling to $H_I\lesssim 10^9\text{ GeV}$ if we restrict $\bar{\theta}_{i,\rm min}=10^{-2}$. We note that such low values of $H_I$ can be achieved in, e.g., Kachru-Kallosh-Linde-Trivedi or brane inflation~\cite{Kachru03,Inflationaris}. $H_I<10^9\text{ GeV}$ implies unobservably small~\cite{Sigurdson05} tensor modes: $r_T<1.6\times 10^{-11}$.

For $\fqcd\lesssim 10^{11}\text{ GeV}$ it is possible to have \UPQ~broken after inflation while satisfying the ULA isocurvature constraint. In this case, the QCD axion suffers from constraints from domain walls~\cite{Sikivie1982,lazarides1982,choi_kim1982}, and uncertainty on the relic density due to axion string decay. It does, however, avoid all isocurvature constraints. The maximum value of $H_{I,\rm max}\approx 10^{12}\text{ GeV}$ imposed by ULA isocurvature constraints thus sets a maximum value of $r_{T,\rm max}\approx 10^{-5}$ in our model. Due to the constraints from domain walls and uncertainty due to string decay, we will not discuss the scenario in which \UPQ~is broken after inflation any further.

Our preferred cosmological model is the following:
\begin{itemize}
\item \UPQ~and \UULA~broken during inflation. Necessary for ULAs. Avoids phase transition relics for QCD axion.
\item $\mula\sim 10^{-22}\text{ eV}$: solution to dSph cusp-core problem if this is the dominant DM component.
\item $\fula\sim 10^{17}\text{ GeV}$: allows for dominant ULA DM.
\item $10^{9}\text{ GeV}<\fqcd<10^{14}\text{ GeV}$: allows for the QCD axion to be up to 10\% of the DM with fine-tuning no worse than $\bar{\theta}_i\gtrsim 10^{-2}$.
\item $H_I<10^9\text{ GeV}$: maximum inflationary Hubble scale allowed by isocurvature if the QCD axion is not fine tuned. 
\end{itemize}

\section{Symmetries and Light Particles}
\label{sec:discrete_symms}

\subsection{Global symmetries and anomalies}

Two axions, the QCD axion, $a$, and the ULA, $\ula$, are introduced by spontaneously breaking the approximate global symmetries \UPQ~and \UULA. The spontaneous breaking is achieved with two SM singlet complex scalar fields, $X_1$ and $X_2$, of which the axions are the angular degrees of freedom. The symmetries \UPQ~and \UULA~are axial symmetries acting as rotations on the angular parts of $X_1$ and $X_2$: hence the axions are pseudoscalars. We work in a SUSY setting where there are two Higgs fields, $H_u$ and $H_d$, which give masses to the up-type, $u$, and down-type, $d$, quarks, $q$, respectively.

If we do not want to consider the QCD anomaly, one may consider only the SM singlet scalars and the global symmetries. But, as soon as the quarks and their PQ charges are taken into account, consideration of the QCD anomaly is inevitable. At a more fundamental level, such as in string theory, it is very hard to make the underlying discrete symmetries blind to quarks. If we construct approximate global symmetries starting from exact discrete symmetries, the SM quarks or some heavy quarks are supposed to carry the discrete charges. So, consideration of the QCD anomaly, and thus the QCD axion as well as a ULA, seems inevitable.

A successful resolution of the strong-$CP$ problem can be achieved if the ULA carries zero QCD anomaly, and we use this to fix the quark charges under the approximate global symmetries. If there are no additional heavy quarks, one may consider the charges shown in Table \ref{tab:NoHq}. These charge assignments ensure that the allowed dimension-4 operators in the Lagrangian respect the approximate \UPQ~and \UULA~symmetries. 

The \UULA~charges are assigned such that there is no \UULA-SU(3)-SU(3) anomaly. There is, however, a \UPQ-SU(3)-SU(3) anomaly, and thus the QCD axion acquires its mass as usual through QCD instantons after the confinement transition. The ULA has no QCD anomaly, but can acquire a mass from higher-order operators in the Lagrangian. We now turn to these.

\begin{table}[!ht]
\begin{center}
\begin{tabular} {c|ccccc|cc}
 & $q_L$& $u_L^c$& $d_L^c$& $H_u$& $H_d$& $X_1$&$X_2$\\[0.2em]\hline
\UPQ & $1$& $1$& $1$& $-2$& $-2$& $2$& $0$\\[0.2em]
\UULA & $1$& $-3$& $1$& $ 2$& $-3$& $0$& $1$
\end{tabular}
\end{center}
\caption{Global charge assignments required for vanishing ULA color anomaly in the model without heavy quarks.}\label{tab:NoHq}
\end{table}

\subsection{Discrete symmetries}

Quantum gravity, thanks to the black hole no-hair theorems, is expected to violate all continuous global symmetries, with the exchange of Planck scale black holes allowing, for example, baryon number violating processes.\footnote{For discussion relating specifically to the QCD axion, see Refs.~\cite{Kamion92,BarrBlock}.} Our axion model involves the introduction of two global symmetries, \UPQ~and \UULA, and both of these symmetries will be broken by Planck-suppressed higher-dimensional operators, inducing effective masses for the axions at low energy. Discrete global symmetries, on the other hand, are allowed by quantum gravity (consider e.g. a $\Z_N$ orbifold compactification of string theory). By imposing an exact discrete symmetry we can control the order of higher-dimensional operators appearing in the effective theory breaking that global symmetry, and thus gain some control over the induced axion masses.

\begin{table}[h!]
\begin{center}
\begin{tabular} {c|ccccc|cc}
 & $q_L$& $u_L^c$& $d_L^c$& $H_u$& $H_d$& $X_1$&$X_2$\\[0.2em]\hline
$\Z_4$ & $1$& $-3$& $1$& $ 2$& $-3$& $0$& $-3$ 
\end{tabular}
\end{center}
\caption{$\Z_4$ example, allowing the approximate global charges of Table~\ref{tab:NoHq}.}\label{tab:Z4}
\end{table}
To have the global charges of Table \ref{tab:NoHq}, we may consider a discrete symmetry $\Z_4$ with the quantum numbers given in Table \ref{tab:Z4}. The discrete charges of Table   \ref{tab:Z4} allow the following terms at lowest order in the low-energy action:
\begin{eqnarray}
&&q_Lu^c_L H_u,\label{eq:tmass}\\[0.3em]
 &&q_Ld^c_L H_d\frac{X_2}{M_{\rm UV}},\label{eq:bmass}\\[0.3em]
&& \frac{H_uH_d X_1^2 X_2}{M_{\rm UV}^2},\label{eq:mu}
\end{eqnarray}
where $M_{\rm UV}$ is a UV mass scale (e.g. the string or the Planck scale).
The term (\ref{eq:tmass}) gives $t$ quark an electroweak scale mass, and the term (\ref{eq:bmass}) gives mass to $b$ quark. For the $b$ quark mass near the electroweak scale the VEV of $X_2$ must be near $M_{\rm UV}$. The term (\ref{eq:mu}) is the one giving the $\mu$-term in SUSY models~\cite{KimNilles84,Casas93}, and defines the \UULA~symmetry. 

The additional term $H_uH_d X_2$ is also allowed by the $\Z_4$ discrete symmetry. However, this spoils a solution of the $\mu$-problem, and must be forbidden. By introducing an additional discrete group $\Z_3$ as in Table \ref{tab:Z43}, we can forbid the unwanted term $H_uH_d X_2$ while keeping (\ref{eq:tmass},\ref{eq:bmass},\ref{eq:mu}). In the last row of Table \ref{tab:Z43}, the combined effect, \ie the $\Z_{12}$ charges are shown.
\begin{table}[!ht]
\begin{center}
\begin{tabular} {c|ccccc|cc}
 & $q_L$& $u_L^c$& $d_L^c$& $H_u$& $H_d$& $X_1$&$X_2$\\[0.2em]\hline
$\Z_4$ & $1$& $-3$& $1$& $ 2$& $-3$& $0$& $-3$ \\[0.2em] 
$\Z_3$ & $2$& $0$& $0$& $ 1$& $1$& $2$& $0$  \\[0.2em] \hline
$\Z_{12}$ & $5$& $9$& $9$& $10$& $1$& $8$& $9$ 
\end{tabular}
\end{center}
\caption{$\Z_4\times \Z_3$ example, where $\Z_3$ is used to ensure a solution to the $\mu$ problem.}\label{tab:Z43}
\end{table}

\subsection{Ultralight axions}

\subsubsection{The simplest model}

For simplicity of discussion, we work in a SUSY framework. The superpotential $W$ gives the information on discrete and global symmetries. As an illustration, consider the global and discrete charges presented in Tables \ref{tab:NoHq} and \ref{tab:Z43}. Obviously, the approximate global symmetries of Table \ref{tab:NoHq} are broken if we consider all the terms in $W$ allowed by the discrete symmetries. \UPQ~is in addition broken by the non-vanishing color anomaly. 

We take the VEV of $X_1$  to fix the decay constant of the QCD axion: $\fqcd/\sqrt{2}= \langle X_1\rangle$. This is a free parameter, which we take to be around $10^{11\,}\gev$. \UULA~is broken by
\begin{eqnarray}
 && \frac{ H_uH_d X_1^2X_2^5  }{M_{\rm UV}^6}.\label{eqn:simple_breaking}
\end{eqnarray}
Eq.~\eqref{eqn:simple_breaking} respects the discrete symmetries of Table \ref{tab:Z43} and the \UPQ~symmetry, but breaks the \UULA~symmetry of Table \ref{tab:NoHq}. When SUSY is broken, this term generates the scale, $V=\fula^2m_{\rm ULA}^2$, of the ULA potential. Taking $\langle X_2 \rangle=\fula/\sqrt{2}=\alpha M_{\rm UV}$ for some constant $\alpha$ we have:
\be
V=\alpha^5s_\beta c_\beta\frac{ m_{3/2}v_{ew}^2\fqcd^2}{M_{\rm UV}}\, ,
\ee
where $s_\beta\equiv \sin\beta = v_u/v_{ew}$ and $c_\beta\equiv\cos\beta=v_d/v_{ew}$ (as is standard in two-Higgs SUSY models). The ULA mass is given by
\begin{widetext}
\be
m_{\rm ULA}=(2\times 10^{-4}\text{ eV})\,\alpha^{3/2}\sqrt{s_\beta c_\beta} \left(\frac{m_{3/2}}{\text{TeV}} \right)^{1/2}\left(\frac{M_{pl}}{M_{\rm UV}} \right)^{3/2}\left(\frac{\fqcd}{10^{11}\text{ GeV}} \right)\left(\frac{v_{ew}}{246\gev} \right) \, .
\label{eqn:mass_simplest}
\ee
\end{widetext}
Setting $\sqrt{s_\beta c_\beta}\approx 1$ and taking all non-axion parameters at the fiducial values given in Eq.~\eqref{eqn:mass_simplest} gives:
\be
\frac{m_{\rm ULA}}{10^{-22}\text{ eV}}\approx 4.25\times 10^{15}\left(\frac{\fula}{10^{17}\text{ GeV}}\right)^{3/2}\left(\frac{\fqcd}{10^{11}\text{ GeV}}\right) \, .
\ee
It is clear that this simplest model cannot produce a cosmology matching our needs: either $\fula$ is too low, or $m_{\rm ULA}$ is too large.
 
\subsubsection{More general discrete charges}

Consider a variation of our simplest model, generalizng the discrete charges we allow our fields to carry, while keeping $\langle X_2\rangle=\fula/\sqrt{2}$, and the global charges of Table~\ref{tab:NoHq}. Take $\Z_N$ with a slight modification of Table \ref{tab:Z4} using arbitrary charges $m$ and $n$ for $X_2$ and $H_d$ respectively, which is shown in Table~\ref{tab:arbZN}.

\begin{table}[!ht]
\begin{center}
\begin{tabular} {c|ccccc|cc}
 & $q_L$& $u_L^c$& $d_L^c$& $H_u$& $H_d$& $X_1$&$X_2$\\[0.2em]\hline
$\Z_N$ & $1$& $-3$& $1$& $ 2$& $n$& $0$& $m$ 
\end{tabular}
\end{center}
\caption{More general $\Z_N$ charges.}\label{tab:arbZN}
\end{table}

We must allow the $t$ quark mass by
\begin{eqnarray}
&&q_Lt^c_L H_u,\label{eq:tmass1} 
\end{eqnarray}
and $b$ quark mass by
  \begin{eqnarray}
 &&q_Ld^c_L H_d\frac{X_2}{M_{\rm UV}} \label{eq:bmass1} \, .
\end{eqnarray}
$\Z_N$ invariance then constrains
  \begin{eqnarray}
 2+n+m=0~{\rm mod~}N\, ,\label{eqn:bmass_modN} 
\end{eqnarray}
\UULA~breaking occurs via the operator,
\be
\frac{1}{M_{\rm UV}^{p+1}}H_uH_dX_1^2 X_2^p \, ,
\label{eqn:ula_break_arbZN}
\ee
for which $\Z_N$ invariance implies 
\be
2+n+mp=0~{\rm mod~}N\, .
\label{eqn:ula_modN}
\ee 

After SUSY breaking the scale of the ULA potential is given as before by:
\be
V=\fula^2m_{\rm ULA}^2=\alpha^ps_\beta c_\beta \frac{m_{3/2}v_{ew}^2\fqcd^2}{M_{\rm UV}} \, ,
\ee
and the mass is
\begin{widetext}
\be
m_{\rm ULA} = (2\times 10^{-4}\text{ eV})\,\alpha^{p/2-1}\sqrt{s_\beta c_\beta} \left(\frac{m_{3/2}}{\text{TeV}} \right)^{1/2}\left(\frac{M_{pl}}{M_{\rm UV}} \right)^{3/2}\left(\frac{\fqcd}{10^{11}\text{ GeV}} \right)\left(\frac{v_{ew}}{246\gev} \right) \, .
\ee
\end{widetext}
Fixing the fiducial values of the non-axion parameters this gives:
\be
\frac{m_{\rm ULA}}{10^{-22}\text{ eV}}\approx 10^{19.69-0.77 p} \left(\frac{\fula}{10^{17}\text{ GeV}}\right)^{p/2-1}\left( \frac{\fqcd}{10^{11}\text{ GeV}}\right)\, .
\label{eqn:fula_mula_zn_general}
\ee
We now have a class of models defined by the value of $p$. 

The system of Eqs.~(\ref{eqn:bmass_modN},\ref{eqn:ula_modN}) is simple to solve for the integers $\{m,n,p\}$. Each solution is characterized by three free integers, $\{\nu_1,\nu_2,\nu_3\}$, since each of $\{m,n,p\}$ is fixed up to addition of some multiple of $N$. We require non-trivial solutions with $m\neq 0$. The most important part of the solution for our purposes is the value of $p=q+N\nu_3$. The term with $p=1$ is \UULA~invariant. This is the SUSY $\mu$-term~\cite{KimNilles84}, and we must allow it. Therefore we require $q=1$, and the lowest order \UULA~breaking term occurs when $\nu_3=1$.  Eq.~\eqref{eqn:bmass_modN} implies there is always a non-trivial solution with $q=1$, and therefore by fixing $N$ we can easily obtain any $p=1+N$.

Let's consider one explicit example. Take the model with 
\be
(N=24,m=8,n=14)\Rightarrow\,p=25\,.
\ee
This gives $m_{\rm ULA}=2.8\times 10^{-22}\text{ eV}$ with $\fula=10^{17}\text{ GeV}$ and $\fqcd=10^{11}\text{ GeV}$.

\begin{figure*}[!t]
\includegraphics[width=0.6\textwidth]{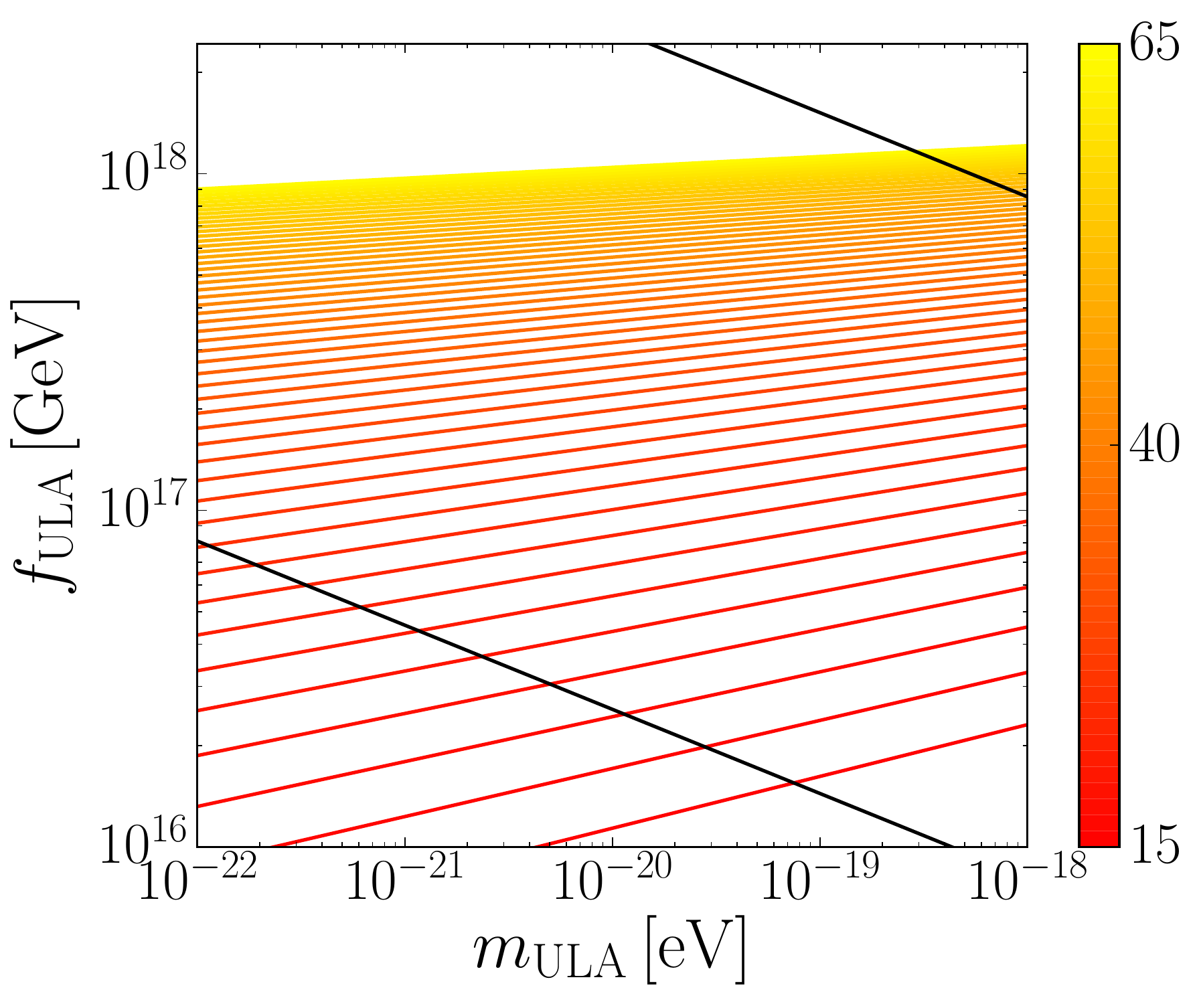}
\caption{ULA models generated by the generalized discrete symmetry model presented in Table~\ref{tab:arbZN}, labelled by the integer parameter $p$ of Eq.~\eqref{eqn:ula_break_arbZN}. The given values of $p$ are most easily achieved for a discrete symmetry $\Z_N$ with $N=p-1$. The demand that the ULA make up $>90\%$ of the DM with fine tuning no greater than $10^{-2}$ constrains the desired models to live between the solid black lines.}
\label{fig:discrete_model_constraints}
\end{figure*}

Fig.~\ref{fig:discrete_model_constraints} shows the space of models generated in this manner, for $p\in[15,65]$, along with the constraints from the DM abundance. We fix all the non-axion parameters to their fiducial values. In addition we fix $\fqcd=10^{11}\text{ GeV}$, which allows the QCD axion to make up anywhere between $\ll 1\%$ and 10\% of the DM with $10^{-2}\lesssim\theta_{i,\rm QCD}\lesssim 0.6$. We impose the constraint that the ULA make up $>90\%$ of the DM with $10^{-2}<\theta_{i,\rm ULA}<1$, which limits the range of allowed $\fula$ for a given $m_{\rm ULA}$. We are able to generate acceptable models across the entire desired range of ULA masses. Interestingly, in this mass range our models seem to display a maximum value of $\fula\sim  10^{18}\text{ GeV}$ consistent with $\langle X_2\rangle<M_{pl}$.

\section{Couplings to the Standard Model}
\label{sec:couplings}

Having built our desired two-axion models, we now compute the couplings of $a$ and $\phi$ to the SM. For the axion and ULA, we define the couplings as
\begin{widetext}
\dis{
{\cal L} \ni &~~ e^2c_{a\gamma\gamma}\frac{a}{32\pi^2 f_a}F_{\rm em,\mu\nu} \tilde{F}_{\rm em}^{\mu\nu} +\bar{c}_{1,ae}\frac{\partial^\mu a}{f_a}\,\bar{e}\gamma_\mu \gamma_5 e + \frac{\partial^\mu a}{f_a}\left(\bar{c}_{1,au}\,\bar{u}\gamma_\mu \gamma_5 u+\bar{c}_{1,ad}\,\bar{d}\gamma_\mu \gamma_5 d\right)\\
 & +e^2 c_{\ula\gamma\gamma}\frac{\ula}{32\pi^2 \fula}F_{\rm em,\mu\nu} \tilde{F}_{\rm em}^{\mu\nu} +\bar{c}_{1,\ula e}\frac{\partial^\mu \ula}{\fula}\,\bar{e}\gamma_\mu \gamma_5 e + \frac{\partial^\mu \ula}{\fula}\left(\bar{c}_{1,\ula u}\,\bar{u}\gamma_\mu \gamma_5 u+\bar{c}_{1,\ula d}\,\bar{d}\gamma_\mu \gamma_5 d\right)
 }
 \end{widetext}
 where the derivative couplings preserve the PQ and ULA symmetries. Using equations of motion, we consider the following couplings,
 \begin{widetext}
 \dis{
{\cal L}\to &~~ e^2 c_{a\gamma\gamma}\frac{a}{32\pi^2 f_a}F_{\rm em,\mu\nu} \tilde{F}_{\rm em}^{\mu\nu} +c_{aee}\frac{m_e }{f_a}\,\bar{e}i\gamma_5  e a + \frac{m_N}{f_a}\left(C_{app}\bar{p}i \gamma_5 p+C_{ann}\bar{n}i \gamma_5 n\right)\,a\\
&+e^2c_{\ula\gamma\gamma}\frac{\ula}{32\pi^2 \fula}F_{\rm em,\mu\nu} \tilde{F}_{\rm em}^{\mu\nu} +c_{\ula ee}\frac{m_e}{\fula}\,\bar{e}i \gamma_5 e\,a + \frac{m_N}{\fula}\left(C_{\ula pp}\,\bar{p}i \gamma_5 p+C_{\ula nn}\,\bar{n}i\gamma_5 n\right)\ula
}
\end{widetext}
where $m_N$ is the nucleon mass, $m_p\approx  m_n\approx m_N$.

For example, from the action $ \int d^4x {\cal L}$ we may consider $ {\cal L}\ni  \frac{\partial^\mu a}{f_a}\,\bar{e}\gamma_\mu \gamma_5 e$, neglecting the surface term
\begin{widetext}
\begin{eqnarray}
&-\frac{a}{f_a}\, \bar{e}\, [\gamma^\mu\partial_\mu e]  +\bar{e}(i\gamma^\mu\partial_\mu )e-m_e\bar{e}\, e
=-\frac{a}{f_a}\, \bar{e}\,  [\gamma^\mu\partial_\mu (e_L+e_R)]~(  {\rm with~} \gamma^\mu\partial_\mu =-im_e)\nonumber\\
&\to a\frac{im_e}{f_a}\, \bar{e}\,(\Gamma_L e_L -\Gamma_R e_R)=(\Gamma_L   -\Gamma_R) \frac{m_e}{f_a}\, \bar{e}i\gamma_5e\, a\label{eq:aeecoup}
\end{eqnarray}
\end{widetext}
where $e_{L,R}=\frac{1\pm\gamma_5}{2}e$. In the last part of (\ref{eq:aeecoup}), we choose the L-chiral representation.  From the \UULA~transformation of $e$, $e_{L,R}\to e^{i\theta \Gamma_{L,R}}e_{L,R}$ with $\theta=\ula/f_{\rm ULA}$, thus we obtain
\begin{eqnarray}
{\cal L}\to (\Gamma_L-\Gamma_R) \frac{m_e}{ f_{\rm ULA}}\,\bar{e}\,i\gamma_5 e\,\ula,
\end{eqnarray}
where $\Gamma_L$ ($\Gamma_R$) is the \UULA~charge of $\ell_L\,(e_R)$ (\ie for $e_R$ it is minus that of $e_L^c$) in the tables with \UULA~quantum number $+1$. Thus, whether $H_d$ or $H_u$ is responsible for the electron mass, we obtain in both cases $\Gamma_L-\Gamma_R=-2$ from Table \ref{tab:NoHq}. On the other hand, if we use the model with heavy quarks, \ie for Table \ref{tab:Hq} in Appendix~\ref{appendix:heavy_q}, we obtain $\Gamma_L-\Gamma_R=-3$. There is a freedom in assigning the electron-ULA coupling at this bottom-up approach. 

For the axion and ULA couplings to neutrons and protons, we  borrow the results from \cite{Kaplan85,ChangS93,KimRMP}.  We use the global quantum numbers presented in Table ~\ref{tab:NoHq}. Note that the phase of $X_1$ is the QCD axion and the phase of $X_2$ is the ULA. Since the ULA charge of $X_2$ is 1, we use the quark ULA charges  those of Table  \ref{tab:NoHq}. On the other hand, the PQ charge of $X_1$ is 2 and hence we use the quark PQ charges the halves of those given in Table  \ref{tab:NoHq}. 
 
 \subsection{QCD axion couplings}\label{subsub1} 
 
 From Table \ref{tab:NoHq}, the electron quantum numbers are read and we obtain $-  m_e/f_a$ for the $aee$ coupling.
For Table \ref{tab:NoHq}, we choose $\delta_{H_u}=-1$ and $\delta_{H_d}=-1$ in Eqs. (61,62) of  \cite{KimRMP},
\dis{
&\bar{c}_1^u= \frac{1}{2(1+Z)}+\frac{v_d^2}{2v_{ew}^2}\simeq -\frac{1}{6}-\frac12 s_\beta^2,\\
&\bar{c}_1^d= \frac{Z}{2(1+Z)}+\frac{v_u^2}{2v_{ew}^2}\simeq -\frac{1}{3}-\frac12 s_\beta^2
}
where $v_{ew}^2=v_u^2+v_d^2, s^2_\beta=v_u^2/v_{ew}^2,  c^2_\beta=v_d^2/v_{ew}^2$, and $Z=m_u/m_d\simeq 0.5$. Now, the nucleon couplings are given by
\begin{widetext}
\dis{
&C_{app} =\bar{c}_1^u\,F+\frac{\bar{c}_1^u-2\bar{c}_1^d}{3}D+\frac{\bar{c}_1^u+\bar{c}_1^d}{6}S
=\left( -\frac{1}{6}-\frac12 s_\beta^2 \right)F +\left(\frac12 +\frac12 s_\beta^2 \right)D+\left(-\frac{1}{12}-\frac{1}{6} s_\beta^2 \right)S ,\\
&C_{ann} =\bar{c}_1^d\,F+\frac{\bar{c}_1^d-2\bar{c}_1^u}{3}D+\frac{\bar{c}_1^u+\bar{c}_1^d}{6}S
=\left( -\frac{1}{3}-\frac12 s_\beta^2 \right)F + \frac16  s_\beta^2\,D+\left(-\frac{1}{12}-\frac{1}{6}  s_\beta^2\right)S .
}
\end{widetext}  
Using the nucleon parameters $F=0.47, D=0.81,$ and $S\simeq  0.13\,(\pm 0.2)$:
\begin{eqnarray}
C_{app}&=0.316+0.148s_\beta^2 \, , \\
C_{ann}&=-0.168-0.122s_\beta^2 \, .
\end{eqnarray}
 
\subsection{ULA couplings}\label{subsub2} 

The value of the coupling $c_{\ula\gamma\gamma}$ depends on the \UULA~charges of the leptons, which we have not yet discussed, as they do not contribute to the color anomaly. For example, if $H_d$ couples to leptons as, $ {\ell}_Le^c_L H_d$, the \UULA~charges of the lepton doublet $\ell_L$ and $e^c_L$ are 1 and 1, respectively, as for $q_L$ and $d^c_L$ of Table \ref{tab:NoHq}. Then, 
\begin{widetext}
\dis{
c_{\ula\gamma\gamma}\propto 1_{\rm from\, q_L}\times  3 \left((\frac23 )^2+(\frac{-1}{3})^2 \right)-3_{\rm from\, u_L^c}\times  3 \left((\frac23 )^2 \right)+1_{\rm from\, d_L^c}\times  3 \left( (\frac{-1}{3})^2 \right)\\
+1_{\rm from\, \ell_L}\times    \left((-1)^2 \right)+1_{\rm from\, e_L^c}\times    \left((-1)^2 \right)=0.
}
\end{widetext}
If $H_u$ couples to leptons as, $ {\ell}_Le^c_L \tilde{H}_u $ where $\tilde{H_u}=i\sigma_2 H_u^*$, the \UULA~charges of the lepton doublet $\ell_L$ and $e^c_L$ are again 1 and 1, and we again obtain $c_{\ula\gamma\gamma}= 0$. 

The above is a key result of this paper: \emph{in the simplest models, ULAs do not have the usual two-photon coupling}. In non-minimal models, it is possible to avoid this conclusion by introducing additional heavy quarks charged under \UULA. We discuss such a possibility in the Appendix.

The electron quantum numbers are read off and we obtain $-2 m_e/f_{\rm ULA}$ for the $\ula ee$ coupling. For the $\ula$ coupling to neutrons and protons, we   borrow the results from \cite{Kaplan85,ChangS93} for the case of $c_3=0$ \cite{KimRMP}.
For Table \ref{tab:NoHq}, we choose $\delta_{H_u}=2$ and $\delta_{H_d}=-3$ in Eqs. (61,62) of  \cite{KimRMP},
\dis{
&\bar{c}_1^u= \frac{1}{2(1+Z)}-\frac{v_d^2}{v_{ew}^2}\simeq -\frac{2}{3}+s_\beta^2,\\
&\bar{c}_1^d= \frac{Z}{2(1+Z)}+\frac{3v_u^2}{2v_{ew}^2}\simeq  \frac{1}{6}+1.5 s_\beta^2.
}
Now, the nucleon couplings are given by
\begin{widetext}
\dis{
&C_{\ula pp} =\bar{c}_1^u\,F+\frac{\bar{c}_1^u-2\bar{c}_1^d}{3}D+\frac{\bar{c}_1^u+\bar{c}_1^d}{6}S
=\left(-\frac23 +s_\beta^2\right)F +\left(-\frac13 -\frac23 s_\beta^2\right)D+\left(-\frac{1}{12}+\frac{5}{12} s_\beta^2\right)S , \\
&C_{\ula nn} =\bar{c}_1^d\,F+\frac{\bar{c}_1^d-2\bar{c}_1^u}{3}D+\frac{\bar{c}_1^u+\bar{c}_1^d}{6}S
=\left(\frac16 +\frac32 s_\beta^2\right)F +\left(\frac12 -\frac{1}{6} s_\beta^2\right)D+\left(-\frac{1}{12}+\frac{5}{12} s_\beta^2\right)S,
} 
\end{widetext}
Using again $F=0.47, D=0.81,$ and $S\simeq  0.13\,(\pm 0.2)$:
\begin{eqnarray}
C_{\ula pp}&=-0.594-0.016s_\beta^2 \, , \\
C_{\ula nn}&=0.473+0.624s_\beta^2 \, .
\end{eqnarray}
 

\section{Detection in the Lab and in Astrophysics}
\label{sec:detection}
\begin{figure*}[!ht]
\includegraphics[width=0.75\textwidth]{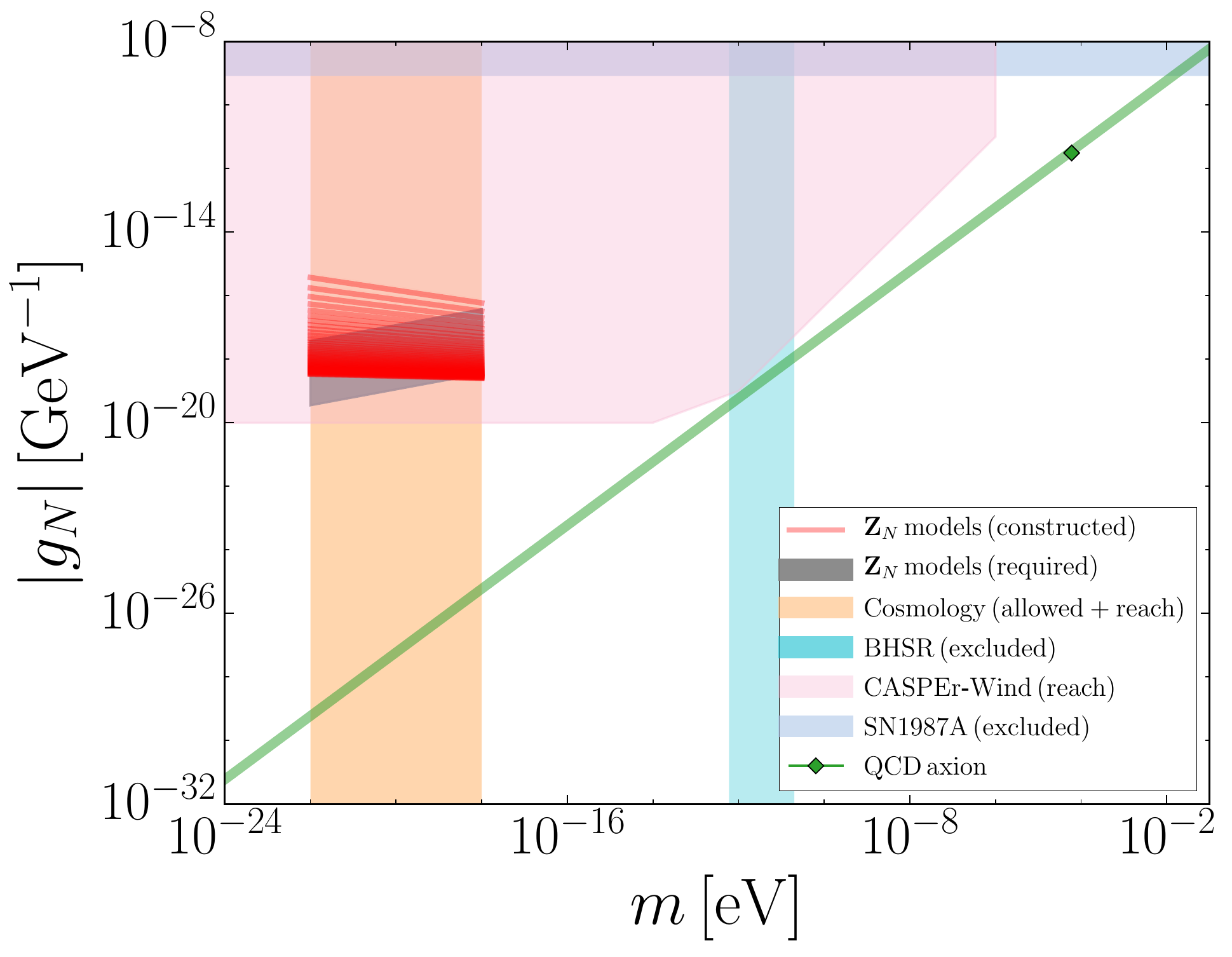}
\caption{Axion parameter space, $(m,g_N)$. The QCD axion paired with the ULA is shown for reference, along with the specific point $\fqcd=10^{11}\text{ GeV}$, which is our reference value. BHSR excludes a range of masses at $2\sigma$ independent of DM abundance and coupling strength~\cite{Arv15}. SN1987A excludes the shaded region with $g_N\gtrsim 8.2\times 10^{-10}$, independent of DM abundance and axion mass~\cite{Raffelt08}. The region both allowed and detectable using cosmology, and relevant to the small-scale crises of CDM is $10^{-22}\text{ eV}\lesssim m\lesssim10^{-18}\text{ eV}$~\cite{Bozek15,SchiveHighZ,MarshPop15,MarshPRD15,SchiveNat14,SchivePRL14}. We show the $\Z_N$ models in this regime only, and also show the target region where $\fula$ allows for the ULAs to be the dominant form of DM without fine tuning. The region accessible to direct detection using the spin precession technique of CASPEr-Wind~\cite{Graham13,casper} is also shown. The cosmologically relevant regime of the $\Z_N$ models lies well within the projected sensitivity of CASPEr-Wind, and is not excluded by any other probes.}
\label{fig:direct_detection}
\end{figure*}

Both the QCD axion and the ULA in our model can be detected directly in the lab by their interactions with the SM. The standard limits from the $\vec{E}\cdot\vec{B}$ coupling (e.g. Ref.~\cite{photon_coupling_review}) apply only to the QCD axion, as the ULA has no such coupling. Furthermore, constraints relying on the QCD axion contributing a large fraction of the DM (e.g. those of ADMX~\cite{admx}) must be scaled appropriately to account for it contributing a sub-dominant DM component in our model. Recently, a whole range of new ideas have emerged for direct detection of axions and ALPs (see e.g. the review in Ref.~\cite{StadnikFlambaum}). We focus here on those aspects novel to the ULA case, where cosmological and astrophysical information is also important.

The most important coupling between ULAs and the SM is the nucleon coupling, and so we consider our models in the plane $(m,g_N)$, shown in Fig.~\ref{fig:direct_detection} (we will discuss the electron coupling briefly later). We display our models as curves in this plane derived from the relationship between $\fula$ and $m_{\rm ULA}$, Eq.~\eqref{eqn:fula_mula_zn_general}, with $12<p<65$, and fixing the reference values of all other parameters, including $\fqcd=10^{11}\text{ GeV}$. For simplicity of presentation, we restrict the ULA mass to the cosmologically relevant regime, but note that our model also works outside this limit. The nucleon coupling is 
\be
g_{NX}=\frac{rC_p}{f_X}+\frac{(1-r)C_n}{f_X}\, ,
\label{eqn:nucleon_coupling}
\ee
where $r$ is the proton-to-neutron ratio in the given medium. As a reference value, we assume a proton-to-neutron ratio taken from the Ref.~\cite{Raffelt08} SN1987A bounds. Our conclusions are not strongly sensitive to this assumption. The thickness of the lines represents the uncertainty based on the unspecified value of $s_\beta$.

The region where ULAs are allowed as a dominant form of DM, yet remain detectable by their imprint on cosmic structure formation is $10^{-22}\text{ eV}\lesssim m \lesssim 10^{-18}\text{ eV}$. In general, ULAs need have no coupling to nucleons, and the region accessible to studies of structure formation are independent of $g_N$. In our model, however, we have $g_N=g_N(\fula)$, and therefore we also show the region where ULAs based on the $\Z_N$ symmetry are able to contribute the dominant form of DM for the given value of $g_N$ (c.f. the region shown in Fig.~\ref{fig:discrete_model_constraints}). The overlap of the model curves, the desired region for our DM model, and the region accessible to studies of structure formation, defines a small window in the $(m,g_N)$ plane.

Black hole superradiance (BHSR)~\cite{Arv15,BritoReview,PaniPRL} provides a constraint on light bosons, independent of coupling and DM abundance. Spinning stellar mass BHs exclude $6\times 10^{-13}\text{ eV}<m<2\times 10^{-11}\text{ eV}$ at $2\sigma$~\cite{Arv15}. These BHSR limits are shown on Fig.~\ref{fig:direct_detection}. Supermassive BHs provide weak constraints on $m\sim 10^{-17}\text{ eV}$, and future studies will improve sensitivity in this regime. The range of masses probed by supermassive BHs is highly complementary to cosmological studies.

The first limits on $g_N$ that we consider are those of Ref.~\cite{Raffelt08} from the cooling of SN1987A, which provides the strongest constraints on the absolute value of the axion-nucleon coupling. The model quoted in Ref.~\cite{Raffelt08} has Helium mass fraction $Y_p=0.3$ and constrains a hadronic axion with $C_p=0.4$, $C_n=0$ to have $f_a\gtrsim 4\times 10^8\text{ GeV}$. This translates into a bound $|g_N|\lesssim 8\times 10^{-10}\text{ GeV}$, which is applicable to all ALPs, regardless of their DM contribution. Our models easily avoid the SN1987A constraint, due to the large values of $\fula$, as shown in Fig.~\ref{fig:direct_detection}.

A direct coupling of axion DM to nucleons induces nuclear spin precession with respect to the DM wind, and this could be detected using nuclear magnetic resonance~\cite{Graham13}. An experiment to detect this effect works on similar principles to the Cosmic Axion Spin Precession Experiment (CASPEr), and has been dubbed ``CASPEr-Wind.'' In contrast, the original CASPEr experiment of Ref.~\cite{casper} has been dubbed ``CASPEr-Electric.'' CASPEr-Electric uses the axion coupling to the neutron electric dipole moment (EDM), and requires an applied electric field in order to detect axions. In our model, the QCD axion has an EDM coupling via $G\tilde{G}$, but the ULA has no such coupling due to the vanishing color anomaly. As the ULA contributes the dominant form of DM, this renders our model largely invisible to CASPEr-Electric. 

CASPEr-Wind searches for axions using the nucleon coupling, and thus requires no applied electric field. Approximate sensitivity curves for CASPEr-Wind are shown in Fig.~\ref{fig:direct_detection}. ULAs oscillate with a very low frequency, and for $m=10^{-22}\text{ eV}$ the spin precession frequency $\nu=m^{-1}\sim 10^{-7}\text{ Hz}$. The corresponding period is $t\sim {\rm months}$. No detailed experimental study has been made of the sensitivity of CASPEr-Wind to such low frequencies, and as such the backgrounds and other issues are unknown. We extrapolate the sensitivity from the lowest mass considered in Ref.~\cite{Graham13}, $m=10^{-14}\text{ eV}$, as a constant. This is likely a conservative extrapolation, assuming that all low frequencies can be constrained simultaneously with a broadband search. Assuming a large $^3$He sample limited by magnetometer noise at low frequencies gives $g_N=10^{-20}\text{ GeV}^{-1}$. This extrapolation easily covers all of our model space. Even the more conservative extrapolation assuming a Xe sample limited by magnetometer and magnetization noise, with $g_N=10^{-18}\text{ GeV}^{-1}$, covers the vast majority of the cosmologically and astrophysically relevant region of the $\Z_N$ ULA models.

\section{Discussion and Conclusions}
\label{sec:conclusions}

We considered models of DM that can simultaneously solve the strong-$CP$ problem and the small-scale crises of CDM, by using two axions with associated approximate global symmetries. In order to solve the small-scale crises of CDM, one axion must be ultralight, with $m_{\rm ULA}\sim 10^{-22}\text{ eV}$, and be the dominant form of DM, requiring a large decay constant, $\fula\sim 10^{17}\text{ GeV}$. The QCD axion must be sub-dominant in the DM and therefore have a relatively small decay constant: our benchmark that requires minimal fine-tuning has $\fqcd\sim 10^{11}\text{ GeV}$. Minimal model uncertainty occurs if both axions are formed by symmetry breaking before inflation. In the simplest case, our model requires low scale inflation, with $H_I\lesssim 10^{12}\text{ GeV}$. 

Axions in general have their masses protected by global symmetries, however these symmetries are broken by quantum gravitational effects. In order to control the quantum corrections to the ULA mass, one can make the global symmetries approximate, descending from some exact , and fundamental (i.e. respected by quantum gravity), discrete symmetry $\Z_N$. Achieving low masses  and large decay constants requires the approximate global symmetries to be of high quality, which requires fixing the order, $N$, of the discrete symmetry. We were able to construct a large family of such models. A benchmark model has $N=24$, and we noted asymptotic behaviour of the models at large $N$. Varying $\fqcd$ shifts the benchmark value of $N$ by approximately one, up or down, per order of magnitude in energy.

In order to maintain a satisfactory solution to the strong-$CP$ problem, we demanded that the QCD axion carry all of the color anomaly. This fixed the charges of the SM fermions under the PQ symmetries and we computed the couplings of the QCD axion and the ULA. We found that, in a minimal model with no additional heavy quarks, the ULA has identically zero coupling to $\vec{E}\cdot\vec{B}$ of electromagnetism, and so is invisible to the standard searches for axion-like particles. The axion-nucleon coupling, however, is non-vanishing, and we explored the astrophysical constraints and sensitivity of direct-detection experiments. 

We found that the cosmologically relevant region of the $\Z_N$ models is well within the direct detection sensitivity of the proposed CASPEr-Wind experiment. This is the main finding of our work. We have constructed a class of models where one can obtain ULAs with their masses protected from quantum corrections by a symmetry. The ULAs will be detectable in the near future by their distinctive imprints on cosmic structure formation compared to CDM. The ULAs are also sufficiently strongly coupled to nucleons to be detectable by nuclear spin-precession relative to the DM wind using nuclear magnetic resonance.

We have so far not considered the fifth-force mediated by the axion-fermion couplings in our model~\cite{Moody84,Dobrescu06}. The force is a Yukawa-type force, which for the ULA is extremely long range, $\sim 10^3$~AU. However, because axions are pseudoscalars, the force is also spin-dependent. Astrophysical objects are not-spin polarized, and so axions do not mediate an astrophysical fifth-force. Our model is not excluded by standard searches for additional forces, which tend to assume the forces are scalar-mediated.  

Current constraints on pseudoscalar forces are reviewed in e.g. Refs.~\cite{Raffelt08,Graham13,Arvanitaki14}. For the nucleon coupling, the constraint from fifth forces is weaker than the SN1987A constraint we have already considered. For the electron coupling, fifth force constraints are weaker than those from White Dwarf cooling. The new technique proposed by Ref.~\cite{Arvanitaki14} using NMR to search for spin-dependent forces will be able to detect the QCD axion in our model, via the monopole-dipole interaction. However, even in an optimistic case where the ULA has scalar couplings as large as the QCD axion, the ULA in our model will be invisible to this technique due to the large value of $\fula$ necessary for the DM abundance.

Since astrophysics and cosmology does not care about the scalar or pseudoscalar nature of the DM, evidence in structure formation for an ultralight particle might be due to a scalar and not an axion. Scalar masses cannot be so easily protected from quantum corrections using the methods we have presented here. However, just like the case of no coupling to the visible sector whatsoever, it is a logical possibility to have an ultralight boson with scalar and dilaton-like couplings. Therefore, consider the constraints of Ref.~\cite{vanTilburg15}, which searches for ultralight DM using atomic spectroscopy via a dilaton-like coupling~\cite{ArvanitakiDilaton}. The sensitivity of such a search peaks at $m\sim 10^{-22}\text{ eV}$, and is stronger than equivalence principle tests. If astrophysical evidence for $m\sim 10^{-22}\text{ eV}$ were to build, while detection with CASPEr-Wind fails, then an experiment such as Ref.~\cite{vanTilburg15} may still find direct evidence for ultralight particles. Building a satisfactory model with ultralight scalars poses an interesting topic for future work.

\acknowledgments{We are very grateful to Peter Graham for discussing with us the existence of CASPEr-Wind and its low frequency sensitivity. J. E. K. is supported in part by the National Research Foundation (NRF) grant funded by the Korean Government (MEST) (NRF-2015R1D1A1A01058449) and by the IBS (IBS-R017-D1-2014-a00). Research at Perimeter Institute is supported by the Government of Canada through Industry Canada and by the Province of Ontario through the Ministry of Research and Innovation. D. J. E. M. acknowledges the support of a Royal Astronomical Society research fellowship, hosted at King's College London.}

\begin{appendix}

\section{Heavy quarks and the ULA photon coupling}
\label{appendix:heavy_q}

In the minimal QCD-ULA models considered in the main text, the ULA has no coupling to photons, $c_{\ula\gamma\gamma}=0$. This is because the charge assignments required for vanishing color anomaly in turn impose vanishing electromagnetic anomaly. This conclusion can be avoided in a non-minimal model, which we now discuss.

To have a nontrivial coupling, $c_{\ula\gamma\gamma}\neq 0$, we introduce \UULA~charge carrying heavy quarks. In Table \ref{tab:Hq}, we list the charges for an example model with a single heavy quark, $Q$. For the photon coupling, we calculate the SM fermion  contribution and also the heavy quark contribution,
\begin{widetext}
\dis{
c_{\ula\gamma\gamma}&\propto -1_{\rm from\, u_L}\times  3 \left((\frac23 )^2  \right)+3_{\rm from\, u_L^c}\times  3 \left((\frac23 )^2 \right)-1_{\rm from\, d_L}\times  3 \left( (\frac{-1}{3})^2 \right)-2_{\rm from\, d_L^c}\times  3 \left( (\frac{-1}{3})^2 \right)\\
&\quad -1_{\rm from\, \ell_L}\times    \left((-1)^2 \right)-2_{\rm from\, e_L^c}\times    \left((-1)^2 \right) +\frac32_{\rm from\, Q_L}\times  3 \left((e_Q )^2  \right)+\frac32_{\rm from\, Q_L^c}\times  3 \left((e_Q)^2  \right) \\
&=9(e_Q)^2-\frac43.
}
\end{widetext}

\begin{table}
\begin{center}
\begin{tabular} {c|ccccc|cc|cc}
 & $q_L$& $u_L^c$& $d_L^c$& $H_u$& $H_d$& $X_1$&$X_2$ & $Q_L$ &$Q_L^c$ \\[0.2em]\hline
\UPQ & $1$& $1$& $1$& $-2$& $-2$& $2$ & $0$& $-1$&$-1$\\[0.2em]
\UULA & $1$& $-3$& $2$& $ 2$& $-3$& $0$& $1$& $-\frac32$&$-\frac32$
\end{tabular}
\end{center}
\caption{Global symmetry charge assignments in the model with a single heavy quark.}\label{tab:Hq}
\end{table} 

\end{appendix}


\begin{thebibliography}{99}
\def\prp#1#2#3{{\it Phys.\,Rep.} {\bf #1} (#3) #2}
\def\rmp#1#2#3{{\it Rev. Mod. Phys.} {\bf #1} (#3) #2}
\def\npb#1#2#3{{\it Nucl.\,Phys.\,B} {\bf #1} (#3) #2}
\def\plb#1#2#3{{\it Phys.\,Lett.\,B} {\bf #1} (#3) #2}
\def\prd#1#2#3{{\it Phys.\,Rev.\,D} {\bf #1} (#3) #2}
\def\prx#1#2#3{{\it Phys.\,Rev.\,X} {\bf #1} (#3) #2}
\def\anp#1#2#3{{\it Annals.\,Phys.} {\bf #1} (#3) #2}
\def\pr#1#2#3{{\it Phys.\,Rev.} {\bf #1} (#3) #2}
\def\prl#1#2#3{{\it Phys.\,Rev.\,Lett.} {\bf #1} (#3) #2}
\def\err#1#2#3{\ {\bf #1} (#3) #2\,(E)}
\def\jhep#1#2#3{{\it JHEP} {\bf #1} (#3) #2}
\def\jcap#1#2#3{{\it JCAP} {\bf #1} (#3) #2}
\def\zp#1#2#3{{\it Z.\,Phys.} {\bf #1} (#3) #2}
\def\epjc#1#2#3{{\it Euro.\,Phys.\,J.\,C} {\bf #1} (#3) #2}
\def\jpg#1#2#3{{\it J.\,Phys.\,G} {\bf #1} (#3) #2}
\def\ijmpd#1#2#3{{\it Int.\,J.\,Mod.\,Phys.\,D} {\bf #1} (#3) #2}
\def\mpla#1#2#3{{\it Mod.\,Phys.\,Lett.\,A} {\bf #1} (#3) #2}
\def\apj#1#2#3{{\it Astrophys.\,J.} {\bf #1} (#3) #2}
\def\nat#1#2#3{{\it Nature} {\bf #1} (#3) #2}
\def\sjnp#1#2#3{{\it Sov.\,J.\,Nucl.\,Phys.} {\bf #1} (#3) #2}
\def\apj#1#2#3{{\it Astrophys.\,J.} {\bf #1} (#3) #2}
\def\frp#1#2#3{Front.\ Phys.\ {\bf #1} (#3) #2}
\def\mnra#1#2#3{{\it Mon.\,Not.\,Roy.\,Astron.\,Soc.} {\bf #1} (#3) #2}
\def\jetpl#1#2#3{{\it JETP\,Lett.} {\bf #1} (#3) #2}
\def\pthp#1#2#3{{\it Prog.\,Theor.\,Phys.} {\bf #1} (#3) #2}
\def\jkps#1#2#3{{\it J.\,Korean\,Phys.\,Soc.} {\bf #1} (#3) #2}\def\dum#1#2#3{(#3) {\bf #1}, #2}

\def\ibid#1#2#3{{\it ibid.} {\bf #1} (#3) #2}
\def\err#1#2#3{\ {\bf #1} {\bf #1} (#3) #2}
\def\err#1#2#3{\ {\it ibid.\,}{\bf #1} (#3) #2\,(E)}
 
\bibitem{Planck15} P.\,A.\,R. Ade \etal(Planck Collaboration), \emph{Planck 2015 results  XIII: Cosmological parameters}, [arXiv:1502.01589].
      
\bibitem{Kim79} J.\,E. Kim, \emph{Weak interaction singlet and strong CP invariance}, \prl{43}{103}{1979} [doi: 10.1103/PhysRevLett.43.103].

\bibitem{SVZ80}   M.A.
Shifman, A.I. Vainshtein, and V.I. Zakharov, \emph{Can confinement ensure natural CP invariance of strong interactions?},  \npb{166}{493}{1980} [doi: 10.1016/0550-3213(80)90209-6]. 

\bibitem{Zhit80} A.P. Zhitnitsky, \emph{On possible suppression of the axion hadron interactions},  {\it Yad.\,Fiz.} (1980) {\bf 31}: 497   [{\it Sov.\,J.\,Nucl.\,Phys.} (1980) {\bf 31}: 260].

\bibitem{DFS81} 
 M. Dine, W. Fischler and M. Srednicki, \emph{A simple solution to the strong CP problem with a harmless axion}, 
\plb{104}{199}{1981}. 

\bibitem{PQ77}
  R.\,D.~Peccei and H.\,R.~Quinn, \emph{CP conservation in the presence of instantons},  \prl{38}{1440}{1977}.
  
\bibitem{Weinberg78} S.~Weinberg, \emph{A new light boson?}, \prl{40}{223}{1978}.

\bibitem{Wilczek78} F.~Wilczek, \emph{Problem of Strong p and t Invariance in the Presence of Instantons}, \prl{40}{279}{1978}.
   
 \bibitem{Baer15} H. Baer, K-Y. Choi, J.E. Kim, and L. Roszkowski,  \emph{Dark matter production in the early Universe: beyond the thermal WIMP paradigm}, \prp{555}{1}{2014}
[arXiv:1407.0017].

\bibitem{NEDMexp06} C.A. Baker, D.D. Doyle, P. Geltenbort, K. Green, M.G.D. van der Grinten, P.G. Harris, P. Iaydjiev, S.N. Ivanov, D.J.R. May, J.M. Penflebury \etal, \emph{An Improved Experimental Limit on the Electric-Dipole Moment of the Neutron}, \prl{97}{2006}{131801} [hep-ex/0602020].

\bibitem{Khlopov85}  M. Khlopov, B.A. Malomed, and Ia.B. Zeldovich, \emph{Gravitational instability of scalar fields and formation of primordial black holes}, \mnra{215}{575}{1985}

\bibitem{SmallCrisis06} D.H. Weinberg, J.S. Bullock, F. Governato, R. Kuzio de Naray, and A.H.G. Peter, \emph{Cold dark matter: controversies on small scales}, PNAS {\bf 112} no. {\bf 40}, 12249 (2014) [arXiv:1306.0913].

\bibitem{SmallFeedback} J. O\~norbe, M. Boylan-Kolchin, J.S. Bullock, P.F. Hopkins, D. Keres, C-A. Faucher-Gigu\`ere, E. Quataert, and N.M. \emph{Forged in FIRE: cusps, cores, and baryons in low-mass dwarf galaxies}, \mnra{454}{2092}{2015} [arXiv:1502.02036]. 

\bibitem{SmallRev} A. Pontzen anf F. Governato, \emph{Cold dark matter heats up}, \nat{506}{171}{2014} [arXiv:1402.1764].

\bibitem{Bond82} J.R. Bond,  A.S. Szalay, and M.S. Turner,  \emph{Formation of Galaxies in a Gravitino Dominated Universe},  \prl{48}{1636}{1982} [doi: 10.1103/PhysRevLett.48.1636].

\bibitem{Bode2001} P. Bode, J.P. Ostriker, and N. Turok, \emph{Halo formation in warm dark matter models}, \apj{556}{93}{2001} [astro-ph/0010389]. 

\bibitem{MNRAS12} A.V. Macci\`o, S. Paduroiu, D. Anderhalden, A. Schneider, and B. Moore, \emph{Cores in warm dark matter haloes: a Catch 22 problem}, \mnra{424}{1105}{2012} [arXiv:1202.1282].

\bibitem{Hu00} W. Hu, R. Barkana, and A.  Gruzinov, \emph{Cold and fuzzy dark matter}, \prl{85}{1158}{2000} [astro-ph/0003365].

\bibitem{MarshSilk14} D.~J.~E.~Marsh and J.~Silk, \emph{A model for halo formation with axion mixed dark matter}, \mnra{437}{3}{2014} [arXiv:1307.1705 [astro-ph.CO]].

\bibitem{MarshFerr14} D.~J.~E.~Marsh and P.~G.~Ferreira, \emph{Ultra-Light Scalar Fields and the Growth of Structure in the Universe}, \prd{82}{103528}{2010} [arXiv:1009.3501].

\bibitem{Park12} C-G.~Park, J-c.~Hwang, and H.~Noh,  \emph{Axion as a cold dark matter candidate: low-mass case},  
\prd{86}{083535}{2012} [arXiv:1207.3124].

\bibitem{Ruffini69} R.~Ruffini, and S.~Bonazzola, \emph{Systems of selfgravitating particles in general relativity and the concept of an equation of state},\pr{187}{1767}{1969}.

\bibitem{Seidel91} E.~Seidel, and W.-M.~Suen, \emph{Oscillating soliton stars}, \prl{66}{1659}{1991}.

\bibitem{SchiveNat14} H-Y. Schive, T. Chiueh, and T, Broadhurst, \emph{Cosmic Structure as the Quantum Interference of a Coherent Dark Wave}, Nature Phys. {\bf 10} (2014) 496  [arXiv:1406.6586].

\bibitem{SchivePRL14} H-Y.~Schive \etal, \emph{Understanding the Core-Halo Relation of Quantum Wave Dark Matter, $\psi$DM, from 3D Simulations}, \prl{113}{261302}{2014}  [arXiv:1407.7762].

\bibitem{MarshPop15} D.~J.~E. Marsh and A-R.~Pop, \emph{Axion dark matter, solitons, and the cusp-core problem}, \mnra{451}{2479}{2015} [arXiv:1502.03456]. 

\bibitem{Hlozek15} R. Hlozek, D. Grin, D.J.E. Marsh, and P.G. Ferreira, \emph{A search for ultra-light axions using precision cosmological data}, \prd{91}{103512}{2015} [arXiv:1410.2896].

\bibitem{Bozek15} B. Bozek, D.J.E. Marsh, J. Silk, and R. F.G. Wyse, \emph{Galaxy UV-luminosity function and reionization constraints on axion dark matter}, \mnra{450}{209}{2015}[arXiv:1409.3544].

\bibitem{SchiveHighZ} H.-Y.~Schive, T.~Chiueh, T.~Broadhurst and K.-W.~Huang, \emph{Contrasting Galaxy Formation from Quantum Wave Dark Matter, $\psi$DM, with $\Lambda$CDM, using Planck and Hubble Data} [arXiv:1508.04621].

\bibitem{MarshPRD15} D.~J.~E.~Marsh, \emph{Nonlinear hydrodynamics of axion dark matter: Relative velocity effects and quantum forces}, \prd{91}{123520}{2015} [arXiv:1504.00308].

\bibitem{MarshPRD13} D.~J.~E.~Marsh, D.~Grin, R.~Hlozek, and P.~G. Ferreira, \emph{Axiverse cosmology and the energy scale of inflation}, \prd{87}{121701}{2013} [arXiv:1303.3008].

\bibitem{MarshPRL14} D.~J.~E.~Marsh, D.~Grin, R. ~Hlozek, and P.~G. Ferreira, \emph{Tensor interpretation of BICEP2 results severely constrains axion dark matter}, \prl{113}{011801}{2014} [arXiv:1403.4216 [astro-ph.CO]].

\bibitem{Dias14}  A.~G.~Dias, A.~C.~B.~Machado, C.~C.~Nishi, A.~Ringwald, P.~Vaudrevange, \emph{The quest for an intermediate-scale accidental axion and further ALPs}, \jhep{1406}{037}{2014} [arXiv:1403.5760]

\bibitem{Graham13} P.~W.~Graham and S.~Rajendran, \emph{New Observables for Direct Detection of Axion Dark Matter}, \prd{88}{035023}{2013} [arXiv:1306.6088].

\bibitem{casper} D.~Budker, P.~W.~Graham, M.~Ledbetter, S.~Rajendran and A.~O.~Sushkov, \emph{Cosmic Axion Spin Precession Experiment (CASPEr)}, \prx{4}{021030}{2014} [arXiv:1306.6089].

\bibitem{Witten84} E.~Witten, \emph{Some properties of SO(32) superstrings}, \plb{149}{351}{1984}.

\bibitem{Svrcek06} P.~Svrcek, E.~Witten, \emph{Axions in string theory}, \jhep{0606}{051}{2006} [hep-th/0605206].

\bibitem{Arvanitaki10} A. Arvanitaki, S. Dimopoulos, S. Dubovsky, N. Kaloper, and J. March-Russell, \emph{String Axiverse}, \prd{81}{123530}{2010}  [arXiv:0905.4720].

\bibitem{KimPRL13} J.~E.~Kim, \emph{Natural Higgs-flavor-democracy solution of the mu problem of supersymmetry and the QCD axion}, \prl{111}{031801}{2013} [arXiv:1303.1822].

\bibitem{Chiueh14} T.~Chiueh, \emph{Why is the Dark Axion Mass $10^{-22} $ eV?} [arXiv:1409.0380].

\bibitem{Krauss89} L.~M.~Krauss and F. ~Wilczek, \emph{Discrete gauge symmetry in continuum theories}, \prl{62}{1221}{1989}.

\bibitem{Kamion92} M.~Kamionkowski and J.~March-Russel, \emph{Planck-Scale Physics and the Peccei-Quinn Mechanism}, \plb{282}{137}{1992} [hep-th/9202003]. 

\bibitem{BarrBlock} S.~M.~Barr and D.~Seckel, \emph{Planck-scale corrections to axion models}, \prd{46}{539}{1992};
R. Holman, S.~D.~H.~Hsu, T. W. Kephart, E. W. Kolb, R. Watkins, and L. M. Widrow, \emph{Solutions to the strong CP problem in a world with gravity},
    \plb{282}{132}{1992} [hep-ph/9203206].
 S. Ghigna, M. Lusignoli and M. Roncadelli, \emph{Instability of the invisible axion}, \plb{283}{278}{1992};
  B. A. Dobrescu, \emph{The strong CP problem versus Planck scale physics}, \prd{55}{5826}{1997} [hep-ph/9609221]

\bibitem{KimPLB13} J. E. Kim, \emph{Abelian discrete symmetries $\Z_N$ and $Z_{nR}$ from string orbifolds} ,\plb{726}{450}{2013} [arXiv:1308.0344 [hep-th]].

\bibitem{KimNilles14} J.~E. Kim and H.~P. Nilles, \emph{Dark energy from approximate U(1)$_{\rm de}$ symmetry},\plb{730}{53}{2014} [arXiv:1311.0012 [hep-th]].

\bibitem{KimJKPS14} J.~E. Kim, \emph{Modeling the small dark energy scale with a quintessential pseudoscalar boson}, \jkps{64}{795}{2014} [arXiv:1311.4545 [hep-ph]].

\bibitem{Grifols96} J.~A. Grifols, E. Masso, and R. Toldra,  \emph{Gamma-rays from SN1987A due to pseudoscalar conversion}, \prl{77}{2372}{1996} [astro-ph/9606028].

\bibitem{Raffelt96} J.~W. Brockway, E.D. Carlson, and G.G. Raffelt, \emph{SN 1987A Gamma-Ray Limits on the Conversion of Pseudoscalars}, \plb{383}{439}{1996} [astro-ph/9605197].

\bibitem{Lyth92} D.~H. Lyth, \emph{Axions and inflation: Vacuum fluctuations}, \prd{45}{3394}{1992}.

\bibitem{Visinelli09} L. Visinelli and P. Gondolo, \emph{Dark Matter Axions Revisited}, \prd{80}{035024}{2009} [arXiv:0903.4377].

\bibitem{Fox04} P. Fox, A. Pierce, and S.~D. Thomas, \emph{Probing a QCD string axion with precision cosmological measurements},  [arXiv:hep-th/0409059].

\bibitem{Wantz09} O. Wantz and E. P. S. Shellard, \emph{Axion cosmology revisited}, \prd{82}{123508}{2010} [arXiv:0910.1066].

\bibitem{Dienes15} K.~R.~Dienes, J.~Kost, and B.~Thomas, \emph{A tale of two time scales: mixing, mass generation and phase transitions in he early universe}, [arXiv:1509.00470].

\bibitem{BKPlanck15} P.~A.~R. Ade etal (BICEP2, Keck, Planck Collaboration), \emph{A Joint Analysis of BICEP2/Keck Array and Planck Data}, \prl{114}{101301}{2015} [arXiv:1502.00612].

\bibitem{Planck15Inflation} P.\,A.\,R. Ade \etal(Planck Collaboration),\emph{Planck 2015 results  XX: Constraints on Inflation}, [arXiv:1502.02114].

\bibitem{Kachru03} S. Kachru, R. Kallosh, A. Linde, and S.~P. Trivedi, \emph{de Sitter Vacua in String Theory}, \prd{68}{046005}{2003} [hep-th/0301240].

\bibitem{Inflationaris} J. Martin, C. Ringeval and V. Vennin, \emph{Encyclop{\ae}dia Inflationaris}, Phys. Dark Univ. 5, 75 (2014), [arXiv:1303.3787].

\bibitem{Sigurdson05} K. Sigurdson and A. Cooray, \emph{Cosmic 21 cm Delensing of Microwave Background Polarization and the Minimum Detectable Energy Scale of Inflation}, \prl{95}{211303}{2005} [astro-ph/0502549]

\bibitem{Sikivie1982} P. Sikivie, \emph{Of axions, domain walls and the early universe}, \prl{48}{1156}{1982}.

\bibitem{lazarides1982} G. Lazarides and Q. Shafi, \emph{Axion models with no domain wall problem}, \plb{115}{21}{1982}.

\bibitem{choi_kim1982} K. Choi and J. E. Kim, \emph{Domain walls in superstring models}, \prl{55}{2637}{1985}.

\bibitem{PlanckCoParameters} P.\,A.\,R. Ade \etal\,(Planck Collaboration),  \emph{Planck 2013 results. XVI. Cosmological parameters}, Astron. Astrophys. 
{\bf 571} (2014) A16  [arXiv: :1303.5076].

\bibitem{Kaplan85} D.B. Kaplan, \emph{Opening the axion window}, \npb{260}{215}{1985}.
 
\bibitem{ChangS93} S. Chang and  K. Choi, \emph{Hadronic axion window and the big-bang nucleosynthesis}, \plb{316}{51}{1993}.

\bibitem{KimRMP} J.\,E. Kim and G. Carosi, \emph{Axions and the strong CP problem},  \rmp{82}{557}{2010} [arXiv:0807.3125].
   
 
\bibitem{KimNilles84} J.\,E. Kim and H.\,P. Nilles,  \emph{The $\mu$ problem and the strong CP problem},  \plb{138}{150}{1984}.

\bibitem{Casas93} J.~A. Casas and C. Munoz, \emph{A natural solution to the $\mu$ problem}, \plb{306}{288}{1993}.
  


\bibitem{photon_coupling_review} G.~Carosi {\it et al}, \emph{Probing the axion-photon coupling: phenomenological and experimental perspectives. A snowmass white paper}, [arXiv:1309.7035].

\bibitem{admx} S.~J.~Asztalos {\it et al}, (ADMX collaboration) \emph{A SQUID-based microwave cavity search for dark-matter axions}, \prl{104}{041301}{2010} [arXiv:0910.5914].

\bibitem{StadnikFlambaum} Y.~V.~Stadnik and V.~Flambaum, \emph{New Atomic probes for Dark Matter detection: Axions, Axion-like particles and Topological Defects}, Mod. Phys. Lett. A29, 1440007 (2014) [arXiv:1409.2986].

\bibitem{Arv15} A.  Arvanitaki, M. Baryakhtar, and X. Huang, \emph{Discovering the QCD Axion with Black Holes and Gravitational Waves}, \prd{91}{084011}{2015} [arXiv:1411.2263].

\bibitem{BritoReview} R.~Brito, V.~Cardoso and P.~Pani, \emph{Superradiance}, Lect. Notes Phys. 906 (2015) [arXiv:1501.06570].

\bibitem{PaniPRL} P.~Pani, V.~Cardoso, L.~Gualtieri, E.~Berti, and A.~Ishibashi, \emph{Black-Hole Bombs and Photon-Mass Bounds}, \prl{109}{131102}{2012} [arXiv:1209.0465].

\bibitem{Raffelt08} G.G. Raffelt, \emph{Astrophysical Axion Bounds}, in {\it Axions}, edited by M. Kuster etal, Lecture Notes in Physics (Springer-Verlag, Berlin, 2008, Vol. 741) [hep-ph/0611350].
  
\bibitem{Moody84} J.~Moody and F.~Wilczek, \emph{New macroscopic forces?}, \prd{30}{130}{1984}. 

\bibitem{Dobrescu06} B.A. Dobrescu and I. Mocioiu, \emph{Spin-Dependent Macroscopic Forces from New Particle Exchange}, \jhep{0611}{005}{2006} [hep-ph/0605342].

\bibitem{Arvanitaki14} A.~Arvanitaki and A.A. Geraci, \emph{Resonant detection of axion mediated forces with Nuclear Magnetic Resonance}, \prl{113}{161801}{2014} [arXiv:1403.1290].

\bibitem{vanTilburg15} K.~V.~Tilburg, N.~Leefer, L.~Bougas, and D.~Budker, \emph{Search for ultralight scalar dark matter with atomic spectroscopy}, \prl{115}{011802}{2015} [arXiv:1503.06886].
 
 \bibitem{ArvanitakiDilaton} A.~Arvanitaki, J.~Huang, and K.~V.~Tilburg, \emph{Searching for dilaton dark matter with atomic clocks}, \prd{91}{015015}{2015} [arXiv:1405.2925].
 
\end{thebibliography}
\end{document}